\documentclass[intlimits,twoside,a4paper]{article}
\usepackage{amsmath,amssymb}
\usepackage{graphicx}
\usepackage{wrapfig}

\usepackage[cp1251]{inputenc}

\usepackage[eqsecnum]{cmpj3}

\usepackage{bm}

\issue{2018}{21}{2}{23602}
\doinumber{10.5488/CMP.21.23602}

\title{Non-additive symmetric mixtures at selective walls}
\author{A. Patrykiejew}
\address{Department for the Modelling of Physico-Chemical 
Processes, Faculty of Chemistry,\\ Maria Curie-Sk{\l}odowska University, 20031 Lublin, Poland}

\date{Received April 12, 2018, in final form May 19, 2018}

\begin{document}
\newcommand{\mtau}{\mbox{\boldmath$\tau$}}
\newcommand{\mqu}{\mbox{\boldmath$q$}}
\newcommand{\mbr}{\mbox{\boldmath$r$}}
\newcommand{\mbR}{\mbox{\boldmath$R$}}
\newcommand{\egs}{$\varepsilon_{\text{gs}}^{\ast}$ }
\newcommand{\ta}[1]{$T^{\ast}=#1$} 
\newcommand{\egsa}[1]{$\varepsilon_{\text{gs}}^{\ast} = #1$}
\newcommand{\egsAA}{$\varepsilon_{\text{gs},\text{A}}^{\ast}$}
\newcommand{\egsBB}{$\varepsilon_{\text{gs},\text{B}}^{\ast}$}
\newcommand{\myint}[1]{$\int #1(x)dx$}

\maketitle

\begin{abstract}
The results of Monte Carlo simulation of adsorption and wetting behaviour of a highly non-additive symmetric mixture at selective walls is discussed. We have concentrated on the interplay between the  
surface induced demixing in the adsorbed films and the properties of the bulk mixture, which exhibits a closed immiscibility loop. It has been shown that the wetting 
behaviour depends on the 
absolute values of the parameters determining the strengths of interaction between the mixture components and the surface, as well as on their difference. 
In general, an increase of the difference between the adsorption energies of the components leads to a decrease of the wetting temperature.
In the cases when the wetting of non-selective walls occurs at the temperatures above the onset of demixing transition in the bulk, an increasing wall selectivity 
leads to a gradual decrease of the wetting temperature towards the triple point, in which the vapour coexists with the mixed and demixed liquid phases. 
When the wetting temperature at the non-selective wall is located below the onset of the demixing transition in the bulk mixture, an increase of the adsorption 
energy of the selected component causes the developing adsorbed films to demix and to show the reentrant mixing upon  
approaching the bulk coexistence. At the temperatures above the onset of the demixing transition in the bulk, the adsorbed films remain demixed up to the bulk 
coexistence and undergo the first-order wetting transition. A rather unexpected finding has been the observation of a gradual increase of the wetting temperature at highly selective walls.
\pacs 68.08.Bc, 68.35.Rh, 07.05.Tp
\end{abstract}

\section{Introduction} 
\label{INTRODUCTION}

Wetting transitions are among the most important surface-induced phase transitions, and have been studied experimentally,
with the help of various theoretical approaches and by computer simulations \cite{wet1,wet2,wet3,wet4,wet5,wet6,wet7,wet8,wet9,wet10,wet11,wet12,chapter}. 
At present, our understanding of wetting at the interface between one-component fluids and solid surfaces is quite advanced.  
When the pressure approaches the value at which the vapour condenses into the liquid, the adsorbed layer on a solid surface 
may behave differently depending on the properties of the fluid and the solid substrate. When the interaction between the fluid and the solid 
is weak, the film remains thin up to the bulk condensation, and this situation corresponds to a partial wetting. On the other hand, when 
the surface potential becomes sufficiently strong, a uniform film of macroscopic thickness develops. In this case, the
fluid wets the surface completely. It has been 
well established under what conditions a complete wetting occurs, and that the wetting may occur via the first-order or the continuous phase 
transition \cite{wet7,wet10,wet12}. In the case of fluid mixtures in contact with a solid, the situation becomes much more complicated. The mixture may exhibit 
a complete or incomplete mixing over a certain ranges of concentrations and temperatures. Moreover, different strengths of interaction between the mixture components and the 
surface may considerably affect the wetting behaviour  \cite{wetmix2,wetmix3,wetmix4,wetmix5,wetmix6,wetmix7,wetmix8,DitLatz,costas,puri,dias,meija,kumar,pohl,ross,tibus,plech}. 
A large number of parameters that describe the adsorption of mixtures at solid surfaces  was one of the reasons why many authors \cite{wetmix4,APLSSS,kirlik,bucior,woyschon,wil1,wil2,AP-1,AP-2} 
have considered the wetting 
behaviour of a simple model of symmetric mixtures \cite{mix1}. In a symmetric mixture, the interactions between the pairs of the like molecules (AA and BB) 
are the same, while the interactions between the pairs of the unlike particles (AB) are different.
In general, the properties of a symmetric mixture are determined by two parameters. One is the ratio of the interaction strengths of the AB and AA (BB) pairs,
$e= \varepsilon_{\text{AB}}/\varepsilon_{\text{AA}}$, and the second is defined by the ratio of the parameters that define the ranges of the interaction between different pairs,
$s = \sigma_{\text{AB}}/\sigma_{\text{AA}}$. 
 
\begin{figure}[!b]
\centerline{\includegraphics[scale=0.4]{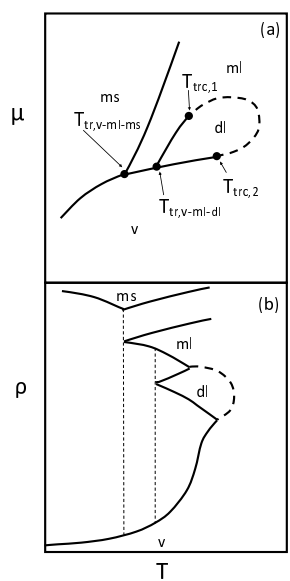}}
\caption{The schematic representation of the bulk phase diagram for a mixture with $e=0.6$ and different values of the parameter $s$, 
between 0.7 and 0.74. Panels~(a) and (b) show ($T^{\ast},\mu^{\ast}$) and ($T^{\ast},\rho)$ projections, respectively.
The regions corresponding to the vapour, the mixed liquid, the demixed liquid and the mixed solid phases are marked by v, ml, dl and ms, 
respectively. $T_{\text{trc},1}$ and $T_{\text{trc},2}$ mark the tricritical points, and $T_{\text{tr,v-ml-ms}}$ and $T_{\text{tr,v-ml-dl}}$ correspond to the different 
triple points. The first-order phase boundaries are given by solid lines, 
and the dashed line corresponds to the $\lambda$-line.}
\label{fig1} 
\end{figure}

The earlier studies were mostly devoted to the wetting of non-selective solid substrates by only energetically non-additive symmetric mixtures, when $e<1$, 
and $s =1$ \cite{wetmix4,APLSSS,kirlik,bucior,woyschon,wil1,wil2,wil3}. The assumption that $s=1$ has made it possible to use different theoretical 
approaches \cite{kirlik,woyschon,wil1,wil2} that cannot be used if $s<1$. In particular, the perturbation and density functional theories require
the knowledge regarding the reference hard sphere systems \cite{chapter}. Though a version of the fundamental measure density functional theory 
for non-additive hard spheres was proposed \cite{matias}, still its application to highly negatively non-additive systems is rather questionable. 
The wetting of non-selective surfaces by energetically ($e<1$) and geometrically ($s\neq 1$) non-additive symmetric mixtures has also been studied, 
but only using the computer simulation methods \cite{AP-1,AP-2}. In a special case of only geometrically non-additive mixtures ($e=1$ and $s\neq 1$), 
which do not phase separate,
the wetting behaviour has been found to strongly depend on the structure of the mixed bulk liquid and solid phases \cite{AP-1}.  
It has been demonstrated that for a given strength of the surface potential, the wetting temperature shows non-monotonous 
changes with $s$, following the changes of the bulk triple point temperature. This occurs even if the wetting temperature is located well above the bulk triple point
and, again, indicates that the short range ordering in the liquid-like films and in the bulk liquids plays an important role in the wetting phenomena.

The study of a wetting behaviour of energetically and geometrically non-additive mixtures, with $e<1$ and $s<1$, at non-selective walls,
has concentrated on the systems that exhibit closed immiscibility loops in the bulk \cite{AP-2}. In such cases, the vapour condenses into 
the mixed liquid at low temperatures, and into the demixed liquid at the temperatures above the triple point, $T_{\text{tr,v-ml-dl}}$, in which the 
vapour coexists with the mixed and demixed liquids. These mixtures show the phase diagrams like the one given in  figure~\ref{fig1}. Since the AB interaction 
is weaker than the AA and BB interaction ($e<1$), the
mixed adsorbed film experiences a stronger attraction to the surface than the demixed film. As a result, the mixed film may wet the surface at the temperatures
below $T_{\text{tr,v-ml-dl}}$. The presence of an attractive surface always increases the 
tendency towards mixing in the film. In the case of negative geometrical non-additivity ($s<1$), the AB pairs are preferred when the density is high, and 
this effect stabilizes the mixed film as well. These two effects lead to the formation of mixed films, when the chemical potential approaches the bulk coexistence, 
even at the temperatures higher than $T_{\text{tr,v-ml-dl}}$. When it happens, only a partial wetting takes place, since the mixed film cannot grow to form 
a macroscopic (mixed) liquid layer. The bulk liquid becomes demixed at $T\geqslant T_{\text{tr,v-ml-dl}}$, and the stable mixed liquid appears only at a higher chemical potential,
at which the demixed and mixed liquids coexist (see figure~\ref{fig1}). However, taking into account that a complete wetting is bound to appear at a certain temperature below 
the temperature at which the vapour-liquid coexistence terminates \cite{wet1}, such systems have been found to show two, i.e., a lower and an upper, wetting transitions~\cite{AP-2}. 

Symmetric mixtures can be treated as a simple model of racemic mixtures \cite{rac1} which consist of R and S isomers. 
The separation of racemic mixtures into the pure R and S
isomers is of great importance in several fields, such as biology and 
drug industry, since the D and L enantiomers very often show different biological activity \cite{rac2}. Thalidomide is probably the best known 
and tragic example, which demonstrated that while one enantiomer of a pharmaceutical can be therapeutic, the other can be very toxic \cite{rac3,rac4}. 
The separation of racemic mixtures can be achieved in a number of ways, and several of them involve adsorption at chiral surfaces \cite{chir1,chir2}. 
For example, kinked-stepped, high Miller index surfaces of metal crystals are chiral and exhibit enantiospecific properties \cite{chir3}. In particular,
the adsorption energies of R and S isomers on such naturally chiral surfaces are different \cite{chir4}. Using the concept of a symmetric mixture, 
it is possible to model an adsorption
on chiral surfaces by assuming that the mixture components interact differently with the solid surface. Although the model is very simple, nevertheless it 
can shed a new light on the interplay between the properties of bulk mixtures and their wetting behaviour.
It is the aim of this paper to elucidate how the 
surface selectivity affects the wetting of highly non-additive symmetric mixtures. In particular, we concentrate on the mixture with the bulk phase diagram 
like the one shown in figure~\ref{fig1}. In particular, we have chosen a mixture which exhibits two wetting transitions, when adsorbed at non-selective walls of a sufficiently 
weak fluid-solid interaction, and attempted to find out whether the same occurs at selective walls. 

\section{The model and simulation method}

The model considered here is the same as that used in  \cite{AP-1,AP-2}. Thus, we have considered the symmetric mixtures consisting of 
the components A and B, interacting via the truncated (12,6) Lennard-Jones potential
\begin{equation}
u_{ij}(r) = \left\{ \begin{array}{ll}
4\varepsilon_{ij}\left[(\sigma_{ij}/r)^{12} - (\sigma_{ij}/r)^{6}\right],   & r \leqslant r_{\text{max}}\,, \\
0,                                                & r > r_{\text{max}}\,,
\end{array}
\right.
\label{eq:01a}
\end{equation}
where $r$ is the distance between a pair of the particles $i$ and $j$. We have assumed that 
$\varepsilon_{\text{AA}}=\varepsilon_{\text{BB}}=\varepsilon$ is the unit of energy, and $\sigma_{\text{AA}}=\sigma_{\text{BB}}=\sigma$ is the unit of length.
Therefore, there are two parameters, $e=\varepsilon_{\text{AB}}/\varepsilon_{\text{AA}}$ and $s=\sigma_{\text{AB}}/\sigma_{\text{AA}}$ that characterize the fluid, 
and we have introduced a reduced temperature, $T^{\ast}=kT/\varepsilon$, 
and the reduced chemical potentials of the species A and B, $\mu^{\ast}_i=\mu_i/\varepsilon$, $i=\text{A,\,B}$. Moreover, the potential energy 
has been expressed in the units of $\varepsilon$. The fluid-fluid interaction potential has been cut at the distance $r_{\text{max}}^{\ast}= 3.0$.
Throughout this paper we have assumed that $\mu^{\ast}_{\text A}=\mu^{\ast}_{\text B}=\mu^{\ast}$.

Unlike in the previous studies \cite{AP-1,AP-2}, the mixture has been assumed to be in contact with a selective smooth wall, 
and the fluid-wall interaction potential, being the function of 
the distance from the wall ($z$) only, has been represented by the following equation:
\begin{equation}
 v_i(z) = \varepsilon_{\text{gs},i}^{\ast}\left[\left(\frac{\sigma}{z}\right)^9 - \left(\frac{\sigma}{z}\right)^3\right].
 \label{eq:01b}
\end{equation}
In the above, $\varepsilon_{\text{gs},i}^{\ast}$ is a measure of the fluid-wall interaction strength (expressed in the units of $\varepsilon$) for the $i$-the 
component (A or B). 

The model has been studied using the Monte Carlo simulation method in the grand canonical ensemble \cite{AP-1,AP-2,FS96,AT87,LB00}. 
Simulations have been carried out using the cells of the size $L_x^{\ast}\times L_y^{\ast}\times L_z^{\ast}=20\times 20\times 60$, with the standard periodic boundary 
conditions in the $x$ and $y$ directions. 
A solid surface has been placed at $z=0$ and the simulation box has been closed from the top, at $z=L_z^{\ast}$, by a reflecting hard wall. The fluid-wall interaction 
potential has been cut at the distance from the surface $z_{\text{max}}^{\ast}=10$.

The quantities recorded included the average numbers of particles A ($N_{\text A}$) and B ($N_{\text B}$), 
the average fluid-fluid potential energy (per particle), $\langle u_{\text{gg}}^{\ast}\rangle$, the average fluid-wall potential energies for the both components 
$\langle u_{\text{gs},k}^{\ast}\rangle$ ($k=\text{A}$ or B) and the number density profiles for each component, $\rho_k(z)$.   
Having the density profiles one can calculate the surface excess densities of the components, $\rho_{\text{ex},k}$, using the following equation
\begin{equation}
\rho_{\text{ex},k} = \frac{1}{S}\int_{0}^{L_z}[\rho_k(z)-\rho_{\text{o},k}]\rd z ,
\label{eq:03}
\end{equation}
 where $S$ is the surface area, $\rho_{\text{o},k}$ is the bulk number density of the $k$-th component, and the total
 excess density $\rho_{\text{ex}} = \rho_{\text{ex,A}}+\rho_{\text{ex,B}}$.
 
 The density profiles allow us to calculate the order parameter profile
 \begin{equation}
  m(z) = \frac{|\rho_{\text A}(z)-\rho_{\text B}(z)|}{\rho_{\text A}(z)+\rho_{\text B}(z)}
  \label{eq:04}
 \end{equation}
being the measure of phase separation in the adsorbed film.

Besides, we have calculated the total order parameter $m$, defined as
\begin{equation}
 m = \frac{|\rho_{\text{ex,B}}-\rho_{\text{ex,A}}|}{\rho_{\text{ex}}}.
 \label{eq:04a}
\end{equation}

In order to study the wetting behaviour we have used the adsorption isotherms calculated at different temperatures and the density profiles recorded
at different temperatures and different values of chemical potential. 
The phase boundaries of the bulk systems have been taken from  \cite{AP-2,AP-NEW}.  
Usually, we have used the runs consisting of $10^8{-}5\cdot 10^9$ Monte Carlo steps. Each Monte Carlo step consisted of randomly chosen attempts to also translate a randomly chosen
particle, to create a new particle in a randomly chosen position, to annihilate a randomly chosen particle or to change the identity of a randomly chosen particle.    
Similar numbers of Monte Carlo steps have been used to equilibrate the system.

Throughout this paper we have considered the mixture with the parameter $e$ equal to 0.6, and the parameter $s$ equal to 0.73.
It has already been demonstrated in our earlier papers \cite{AP-2,AP-NEW} that the bulk phase diagram of this mixture is like the one depicted in  
 figure~\ref{fig1}, and exhibits a closed immiscibility loop. At low temperatures, between the triple point $T^{\ast}_{\text{tr,v-ml-ms}}$, 
in which the vapour coexists with the mixed liquid and with the mixed solid,  and the triple point $T^{\ast}_{\text{tr,v-ml-dl}}$, in which the vapour coexists with the mixed and with demixed liquids,
the vapour condenses into the mixed liquid. Only at the temperatures exceeding the triple point $T^{\ast}_{\text{tr,v-ml-dl}}$, the demixing transition accompanies the vapour condensation.
It should be noted that the demixing is the first order transition at the temperatures between $T^{\ast}_{\text{tr,v-ml-dl}}$ and the tricritical point $T^{\ast}_{\text{trc},1}$, and becomes a 
continuous transition at still higher temperatures. The vapour-liquid coexistence terminates at still another tricritical point, $T^{\ast}_{\text{trc},2}$, which replaces the usual 
vapour-liquid critical point, and is the onset of the continuous demixing transition. 
Thus, the continuous demixing transition occurs along the so-called $\lambda$-line, as shown by the dashed line in figure~\ref{fig1}.

As it has already been stated in the introduction, the wetting behaviour of the mixture considered here, at non-selective walls strongly depends on the strength of the fluid-solid interaction, and
different scenarios are possible. In general, the surface field enhances the mixing in the adsorbed film and since the parameter $e$ is considerably lower than unity, the  relative strength
of the fluid-solid to the fluid-fluid interaction is higher in the mixed film than in the demixed film. One should note that the ratio of densities in the mixed and in the demixed dense thick adsorbed films
is proportional to $s^{-3}$. Therefore, the mixed film is expected to wet the surface 
at lower temperatures than the demixed film. We have found \cite{AP-2} that over a certain range of \egs the wetting transition occurs at the temperature below the triple point 
$T^{\ast}_{\text{tr,v-ml-dl}}$. Then, at $T^{\ast}_{\text{tr,v-ml-dl}}$ and above it, only a partial wetting appears. However, another wetting transition occurs at a
sufficiently high temperature, but below the tricritical point $T^{\ast}_{\text{trc},2}$.

\section{Results and discussion}

The calculations has been carried out for the mixture being in contact with the surfaces characterized by different values of
{\egsAA} and \egsBB. We recall that in the bulk mixture, the vapour condenses into a mixed liquid at the temperatures below 
$T^{\ast}_{\text{tr,v-ml-dl}}\approx 0.92$, and into a demixed liquid at higher temperatures. 

To begin with, we present the results obtained in the limit of low adsorption, i.e., in the Henry's low region, 
when one can assume that the adsorbed molecules do not interact one with another.
In this case, the surface excesses of the A and B components are independent of $s$ and $e$, and depend only on the strengths of the fluid-surface interaction of
the mixture components 
and the temperature, and are given by the following equation
\begin{equation}
 \rho_{\text{ex},k} \sim \rho_{\text{o},k}\int_{z_\text{o}}^{\infty}\big\{\exp[-v_{k}(z)/T^{\ast}]-1\big\}\rd z.
 \label{eq:3.01}
\end{equation}
where $k=\text{A}$ or B. In the above, the integration starts at the Gibbs dividing surface located at $z_\text{o}$ and, of course, the bulk densities of the components A and B are bound to be the same as long as
their chemical potentials are equal. 
Having the surface excesses $\rho_{\text{ex,A}}$ and $\rho_{\text{ex,B}}$,  one can calculate the mole fractions of the components ($x_{\text A}$ and $x_{\text B} = 1-x_{\text A}$) in the adsorbed film, and
the order parameter $m$ [cf. (\ref{eq:04a})].

\begin{figure}[!b]
\centerline{\includegraphics[scale=0.5]{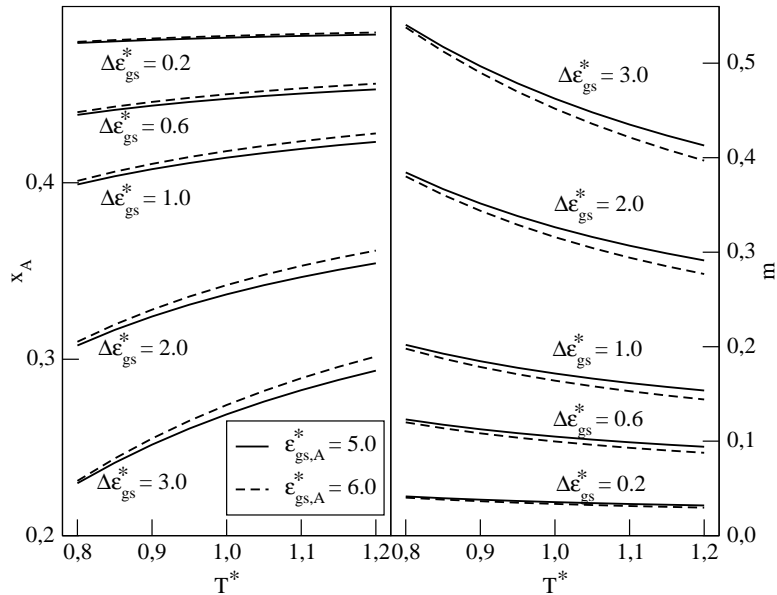}}
\caption{The changes of the mole fraction of the component A (left-hand panel) and the order parameter~$m$ (right-hand panel) versus temperature in the low adsorption limit (in the Henry's region) 
for different values of $\Delta\varepsilon_{\text{gs}}^{\ast} = \text{\egsBB} - \text{\egsAA}$ and two choices of {\egsAA} (given in the figure).} 
\label{fig2}
\end{figure}

Figure~\ref{fig2} shows the changes of the mole fraction of the component A and of the order parameter $m$ with temperature for several choices of $\Delta\varepsilon_{\text{gs}}^{\ast} = \text{\egsAA} -\text{\egsBB}$ and  
for the two values of {\egsAA} equal to 5.0 and 6.0. As expected, an increase of $\Delta\varepsilon_{\text{gs}}^{\ast}$ causes  the component B to be adsorbed more strongly, and  
the enhancement of the B adsorption decreases with the temperature. It is also clear that for the same value of $\Delta\varepsilon_{\text{gs}}^{\ast}$, the enhancement of the B adsorption 
is slightly lower when {\egsAA} is higher. This results from a decrease of the relative strengths of the fluid-surface interaction of the components ($\text{\egsBB}/\text{\egsAA}$) with~{\egsAA}.

When the chemical potential increases, the adsorbed molecules start to interact and the relative amounts of the components change, since the 
magnitudes of the parameters $s$ and $e$ become relevant. 
Different scenarios are possible, depending on the value of {\egsAA}, the magnitude of  $\Delta\varepsilon_{\text{gs}}^{\ast}$, and the location of the wetting temperature 
with respect to the demixing transition in the bulk ($T^{\ast}_{\text{tr,v-ml-dl}}$).

It has been found \cite{AP-2} that when the mixture with $s=0.73$ interacts with the non-selective wall, and {\egsAA} is sufficiently small, 
the wetting occurs only at $T^{\ast}_{\text w}>T^{\ast}_{\text{tr,v-ml-dl}}$. 
In particular, when \egsa{5}, the wetting temperature has been estimated to be
located at $T^{\ast}_{\text w} \approx 1.00$. An increase of \egs to 5.5 shifts the wetting temperature down to about 0.93, i.e., still above the triple point $T^{\ast}_{\text{tr,v-ml-dl}}\approx 0.92$.
On the other hand, when \egs is equal to 6.0, the wetting transition appears already below the triple point $T^{\ast}_{\text{tr,v-ml-dl}}$ at $T^{\ast}_{\text w} = 0.855\pm 0.005$. 
For still higher values of \egsa{6.5} and 7.0, the triple point wetting at $T^{\ast}_{\text w}=T^{\ast}_{\text{tr,v-ml-ms}}$ has been found. 
In such cases, the wetting does not occur below the triple point, 
which is related to the strain induced by the structural mismatch 
between successive solid-like layers \cite{Huse,Gittes}. A pinning of the wetting transition at the triple point (triple point wetting) has been observed 
for several systems, like various gases adsorbed on graphite \cite{Zimmerli,bart,krim1}, on Ag \cite{mig1,mig2}, on Au \cite{krim2,brus}, and on alkaline metals \cite{alk1}. 

In the cases when $T^{\ast}_{\text w}>T^{\ast}_{\text{tr,v-ml-dl}}$, the adsorbed films formed at non-selective walls have been
observed to remain 
mixed even at the temperatures above $T^{\ast}_{\text{tr,v-ml-dl}}$. Only at the temperatures $T^{\ast}\geqslant T^{\ast}_{\text w}$, the adsorbed films have become demixed. 
In the case of a selective wall, 
an increase of $\Delta\varepsilon_{\text{gs}}^{\ast}$ is bound to enhance the demixing tendency in the adsorbed layer. 
Consequently, the wetting temperature has been expected to gradually decrease towards $T^{\ast}_{\text{tr,v-ml-dl}}$, when the wall selectivity increases. 
However, when $\Delta\varepsilon_{\text{gs}}^{\ast}$ becomes large enough,
it may cause a demixing in the entire film up to the bulk coexistence, even at the temperatures below $T^{\ast}_{\text{tr,v-ml-dl}}$. When it happens, 
the demixed film cannot wet the surface, since the bulk liquid remains mixed. This implies that a sort of a triple point wetting at $T^{\ast}_{\text{tr,v-ml-dl}}$ should occur.

On the other hand, when {\egsAA} is sufficiently large, the wetting temperature of the system with $\Delta\varepsilon_{\text{gs}}^{\ast}=0$ (a non-selective wall) occurs below $T^{\ast}_{\text{tr,v-ml-dl}}$,
and the mixed wetting layer coexists with the mixed liquid. When the wall becomes selective with respect to the mixture components, two different cases can be singled out.
In the first, the growing thick film may remain mixed and hence wet the surface. This is expected to happen when  $\Delta\varepsilon_{\text{gs}}^{\ast}$ is sufficiently small, or 
when the demixed film undergoes a reentrant mixing upon  approaching the bulk coexistence. In the second situation, when $\Delta\varepsilon_{\text{gs}}^{\ast}$ is sufficiently large, the film may
remain demixed up to the bulk coexistence, and, therefore, a complete wetting below $T^{\ast}_{\text{tr,v-ml-dl}}$ cannot take place. If this is the case, 
a complete wetting appears only at $T^{\ast}_{\text{tr,v-ml-dl}}$, and a triple point wetting is expected to occur again.

Here, we have discussed the changes in the wetting behaviour of the selected mixture adsorbed at the selective walls characterized by different values of 
{\egsAA} and $\Delta\varepsilon_{\text{gs}}^{\ast}$. 
To this end, we have considered two series of systems, in which the interaction between the component A and the surface is given by {\egsAA} equal to 5.0 and 6.0, respectively, 
while the interaction of the component B with the surface, specified by \egsBB, is varied and is assumed to be higher than \egsAA, i.e.,  $\Delta\varepsilon_{\text{gs}}^{\ast}>0$.

\subsection{Systems with $\text{\egsAA}=5$}

\begin{figure}[!t]
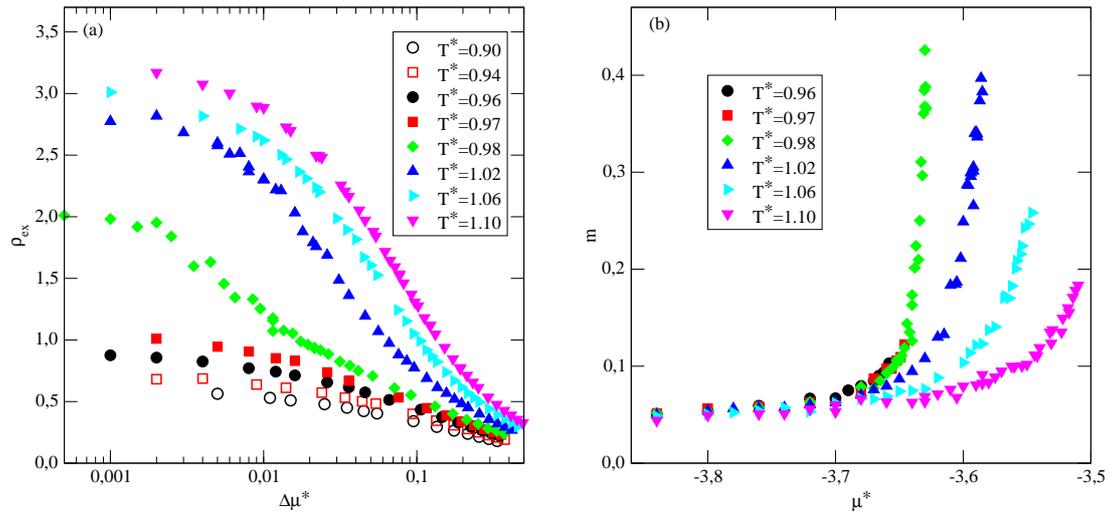

\centerline{\includegraphics[scale=0.4]{Fig03a}\qquad\includegraphics[scale=0.4]{Fig03b}}
\caption{(Colour online) Panel~(a) shows the semi-log plots of the excess adsorption isotherms ($\rho_{\text{ex}}$ versus $\Delta\mu^{\ast}$), while panel~(b) shows the changes of the order parameter $m$ with 
the chemical potential $\mu^{\ast}$ at different temperatures, for the system with $\text{\egsAA}=5.0$ and $\Delta\varepsilon_{\text{gs}}^{\ast} = 0.2$.}
\label{fig3}
\end{figure}

\begin{figure}[!t]
\centerline{\includegraphics[scale=0.5]{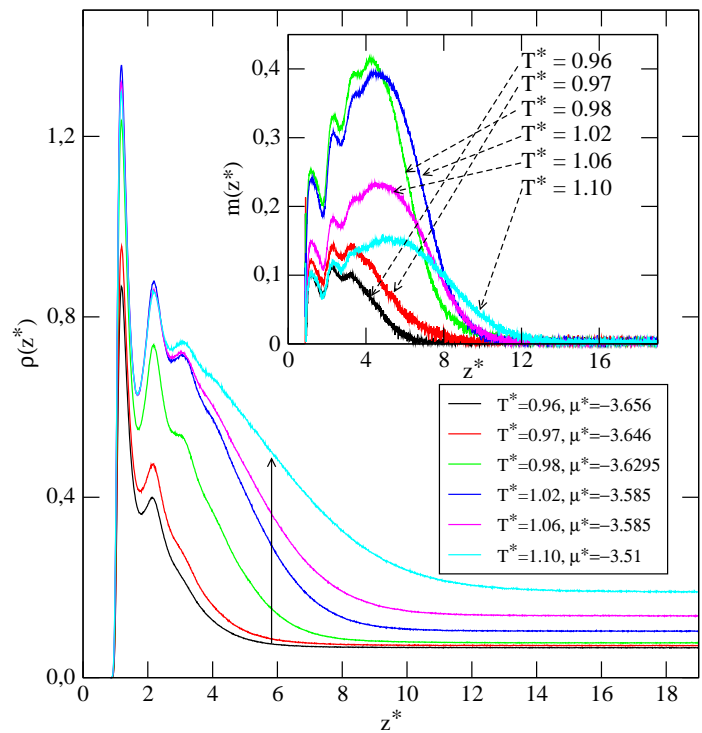}}
\caption{(Colour online) The main figure show the examples of density profiles for the system with $e=0.6$, $s=0.73$, $\text{\egsAA}=5.0$ and $\Delta\varepsilon_{\text{gs}}^{\ast} = 0.2$ at different temperatures and at the
chemical potential values close to the bulk coexistence. The colour key applies to the on-line version of the paper, and the vertical arrow shows the direction of the temperature changes, from the
lowest to the highest.
The inset shows the corresponding order parameter profiles.}
\label{fig4}
\end{figure}

At first, we assume that $\text{\egsAA}=5$ and consider a series of systems with  different $\Delta\varepsilon_{\text{gs}}^{\ast}$ between 0.2 and 1.0.
The panel~(a) of figure~\ref{fig3} presents the semi-log plots of the excess adsorption isotherms ($\rho_{\text{ex}}$ versus $\Delta\mu^{\ast}$), for the system with $\Delta\varepsilon_{\text{gs}}^{\ast} = 0.2$, 
recorded at different temperatures. One readily notes that at $T^{\ast}$ up to 0.97, the total surface excess density stays low, and reaches much lager values at higher temperatures.
In the panel~(b) to figure~\ref{fig3} we present the changes of the order parameter $m$ along the isotherms, which demonstrate that only at the temperatures $T^{\ast}\geqslant 0.98$ the demixing transition
in the film occurs when the chemical potential approaches the bulk coexistence. These results suggest that the wetting occurs at $T^{\ast}_{\text w}\approx 0.98$, i.e., well above
the triple point $T^{\ast}_{\text{tr,v-ml-dl}}$. 
One should note that the degree of demixing in the film, in the vicinity of the bulk coexistence, gradually decreases with temperature. This is caused by a the fact that the bulk liquid also exhibits
a limited phase separation at high temperatures, and by the changes in the film structure with temperature. 
The main panel of figure~\ref{fig4} shows a series of density profiles, recorded close to the bulk coexistence at different temperatures, while the inset shows the corresponding order parameter profiles. 
First we note that at $T^{\ast}<T^{\ast}_{\text w}$, i.e., at $T^{\ast}<0.98$, the film remains thin and the order parameter reaches the magnitude of about 0.1 in the part of the film close to the surface. 
This is about two times lager value than the one obtained in the limit of low adsorption (cf. the right-hand panel of figure~\ref{fig2}). As soon as the temperature reaches 0.98, the film thickness attains larger values and the film 
undergoes a demixing transition. In particular, the parameter $m$ in the first layer jumps to the value of above 0.2 and reaches the values close to 0.4 in the higher layers. 
When, however, the temperature increases further and becomes closer to the tricritical point $T^{\ast}_{\text{trc},2}$, the degree of the film demixing gradually decreases, while the 
interface between the film and the bulk becomes more and more diffused. The demixing in the outer part of the film is somewhat suppressed by the low film density and by the gradually increasing entropic effects.


\begin{figure}[!t]
\centerline{\includegraphics[scale=0.4]{Fig05a}\qquad\includegraphics[scale=0.4]{Fig05b}}
\caption{(Colour online) Panel~(a) shows the semi-log plots of the excess adsorption isotherms ($\rho_{\text{ex}}$ versus $\Delta\mu^{\ast}$), while panel~(b) shows the changes of the order parameter $m$ with 
the chemical potential $\mu^{\ast}$ at different temperatures, for the system with $\text{\egsAA}=5.0$ and $\Delta\varepsilon_{\text{gs}}^{\ast} = 0.8$.}
\label{fig5}
\end{figure}
\begin{figure}[!t]
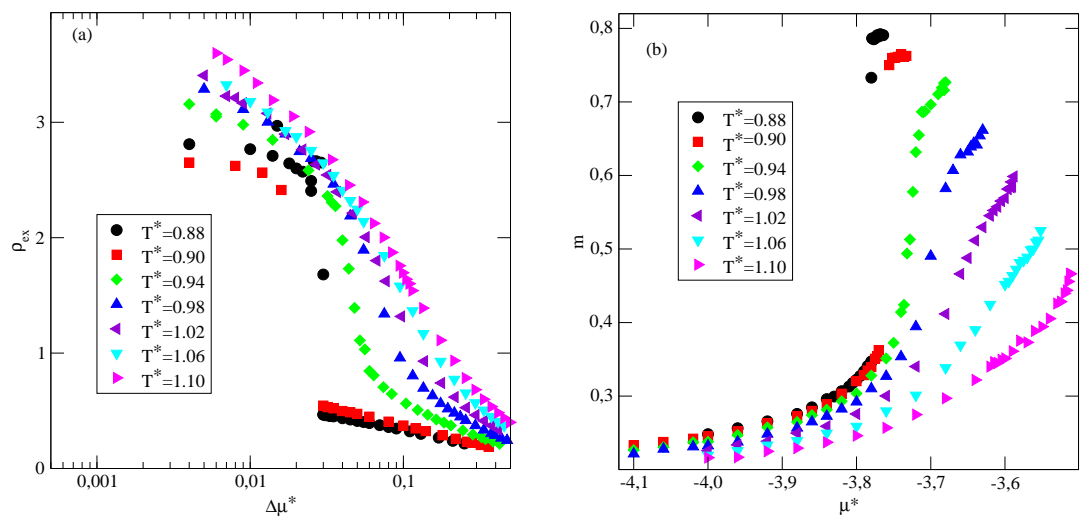

\centerline{\includegraphics[scale=0.4]{Fig06a}\qquad\includegraphics[scale=0.4]{Fig06b}}
\caption{(Colour online) Panel~(a) shows the semi-log plots of the excess adsorption isotherms ($\rho_{\text{ex}}$ versus $\Delta\mu^{\ast}$), while panel~(b) shows the changes of the order parameter $m$ with 
the chemical potential $\mu^{\ast}$ at different temperatures, for the system with $\text{\egsAA}=5.0$ and $\Delta\varepsilon_{\text{gs}}^{\ast} = 1.0$.}
\label{fig6}
\end{figure}

\begin{figure}[!t]
\centerline{\includegraphics[scale=0.4]{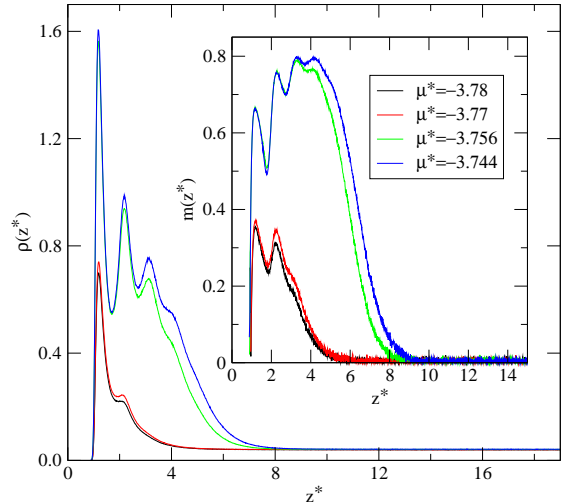}}
\caption{(Colour online) The main figure shows the density profiles for the system with $\text{\egsAA}=5.0$ and $\Delta\varepsilon_{\text{gs}}^{\ast} = 1.0$ at $T^{\ast} = 0.9$ and different values of the chemical potential.
The inset gives the corresponding order parameter profiles.}
\label{fig7}
\end{figure}

When $\Delta\varepsilon_{\text{gs}}^{\ast}$ is increased to $0.4$, and then to 0.6, the behaviour remains qualitatively the same, but the wetting transition occurs at  lower temperatures 
of about 0.95 and 0.92, respectively.  One should note that the triple point at which the vapour coexists with the mixed and demixed liquids is located at the temperature equal to about  $T^{\ast}_{\text{tr,v-ml-dl}}=0.92$. 
Thus, the systems with $\Delta\varepsilon_{\text{gs}}^{\ast}$ up to 0.6 do not 
show the wetting below $T^{\ast}_{\text{tr,v-ml-dl}}$. In the particular case of $\Delta\varepsilon_{\text{gs}}^{\ast}=0.6$, we have found a triple point wetting.
A further increase of $\Delta\varepsilon_{\text{gs}}^{\ast}$ to 0.8 and 1.0 causes a strong preferential adsorption of the component B that leads to the formation of rather thick demixed films already at $T^{\ast}=0.90$ and 0.88,
respectively. The films undergo a sort of layering transition, accompanied by the demixing, but attain only a limited thickness at the bulk coexistence (see figures~\ref{fig5} and \ref{fig6}). 
This suggests that only a partial wetting occurs below $T^{\ast}_{\text{tr,v-ml-dl}}$. Figure~\ref{fig7} gives the examples of the density and of the order parameter profiles for a system 
with $\Delta\varepsilon_{\text{gs}}^{\ast}=1.0$, recorded at $T^{\ast}=0.90$
and at the chemical potentials below and above the layering transition. Below the layering transition point, the film density is rather low, and shows only a limited enrichment in the B component. On the
other hand, a thick film has a considerably larger density and also a considerably higher degree of 
demixing in the entire film. Evidently, the difference between the strengths of the fluid-surface interactions of the components is large enough to trigger the demixing in the film. However, 
when the bulk fluid is mixed, the demixed film cannot grow to reach a macroscopic thickness, implying that only a partial wetting takes place. 
Only when the temperature exceeds  the triple point temperature, $T^{\ast}_{\text{tr,v-ml-dl}}\approx 0.92$,
the demixed film coexists with the demixed bulk liquid at $\mu^{\ast}_\text{o}$, and wets the surface. Therefore,  
in the both cases of $\Delta\varepsilon_{\text{gs}}^{\ast}=0.8$ and 1.0, a  triple point wetting occurs, $T^{\ast}_{\text w} = T^{\ast}_{\text{tr,v-ml-dl}}\approx 0.92$, 
just the same as in the case of $\Delta\varepsilon_{\text{gs}}^{\ast} = 0.6$. 
In all the above considered cases, the wetting is expected to occur via the first-order phase transition, and hence the demixed film reaches a finite thickness at 
the bulk coexistence ($\mu^{\ast}_\text{o}$), where it jumps to infinity. 

\subsection{Systems with $\text{\egsAA}=6.0$}

The second series of calculations have involved the systems with $\text{\egsAA}=6.0$ and different values of $\Delta\varepsilon_{\text{gs}}^{\ast}$, between 0.2 and 1.0. 
We recall that the wetting temperature in the system with $\text{\egsAA}= \text{\egsBB}=6.0$ is located at $T^{\ast}_{\text w}\approx 0.855$ \cite{AP-2}, 
i.e., well below the triple point $T^{\ast}_{\text{tr,v-ml-dl}}$, where the film and the bulk liquid are both mixed.   
In the case of a non-selective wall, the wetting has been observed to be preceded by the prewetting transition, which implies that the wetting 
is the first order transition. It is also known that a complete wetting also occurs at $T^{\ast}\geqslant 0.94$, that is above the triple point $T^{\ast}_{\text{tr,v-ml-dl}}$ (cf.  \cite{AP-2}). 

\begin{figure}[!t]
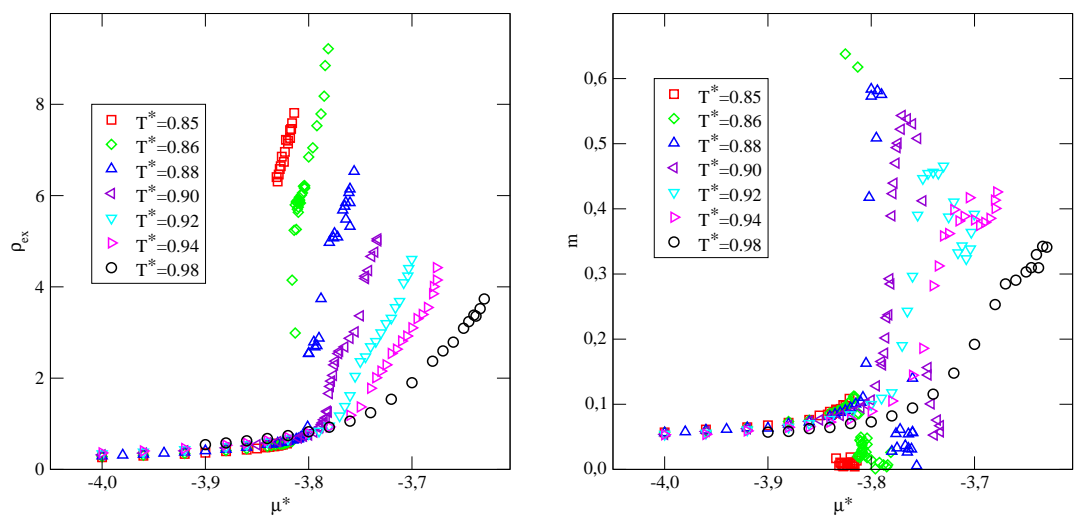

\centerline{\includegraphics[scale=0.4]{Fig08a}\qquad\includegraphics[scale=0.4]{Fig08b}}
\caption{(Colour online) Panels~(a) and (b) show the excess adsorption isotherms ($\rho_{\text{ex}}$ versus $\mu^{\ast}$) and the changes of the order parameter $m$ with 
the chemical potential $\mu^{\ast}$ at different temperatures, respectively, for the system with $\text{\egsAA}=6.0$ and $\Delta\varepsilon_{\text{gs}}^{\ast} = 0.2$.}
\label{fig8}
\end{figure}

In the case of only weakly selective walls, when $\Delta\varepsilon_{\text{gs}}^{\ast}=0.2$, the wetting has also been found to occur below the triple point $T^{\ast}_{\text{tr,v-ml-dl}}$, 
at a slightly lower temperature of $T^{\ast}_{\text w}\approx 0.845$.
The excess adsorption isotherms, shown in figure~\ref{fig8}~(a), demonstrate the presence of the prewetting transition at low temperatures. At $T^{\ast} = 0.85$, this transition leads to 
the formation of a thick mixed film. However, at the temperatures between 0.86 and $T^{\ast}_{\text{tr,v-ml-dl}}$, we have found a different behaviour. Namely, the film undergoes a transition 
between a thin mixed layer and a considerably thicker demixed layer. However, the demixed film undergoes another transition and forms a still thicker layer. The transition is accompanied 
by a reentrant mixing of the film upon approaching the bulk coexistence. This is
demonstrated by the changes of the order parameter $m$ along the isotherms, shown in figure~\ref{fig8}~(b). One should note that the reentrant mixing occurs only at the temperatures below  $T^{\ast}_{\text{tr,v-ml-dl}}$.

\begin{figure}[!t]
\centerline{\includegraphics[scale=0.4]{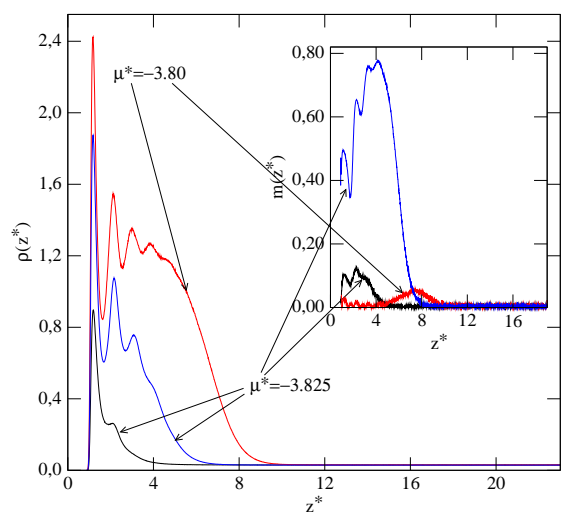}}
\caption{(Colour online) The main figure shows the density profiles for the system with $\text{\egsAA}=6.0$ and $\Delta\varepsilon_{\text{gs}}^{\ast} = 0.2$ at $T^{\ast} = 0.86$ and different values of the chemical potential.
The inset gives the corresponding order parameter profiles. }
\label{fig9}
\end{figure}

Figure~\ref{fig9} presents the density and the order parameter  profiles recorded at different points along the adsorption isotherm at \ta{0.86}. These results demonstrate that a very thin film 
shows only a small enrichment in component B, resulting from the difference in the adsorption energies. Then, a transition to a considerably thicker film occurs and leads to 
the formation of a highly demixed layer. The density of this demixed layer is still not very high and the difference between {\egsAA} and \egsBB, together with a rather weak interaction between the pairs of
unlike particles, trigger a demixing transition in the film. However, when the chemical potential becomes closer to the bulk coexistence, another transition occurs, and a still thicker film of higher
density is formed. This film is mixed, since the packing effects start to dominate. 

\begin{figure}[!t]
\centerline{\includegraphics[scale=0.4]{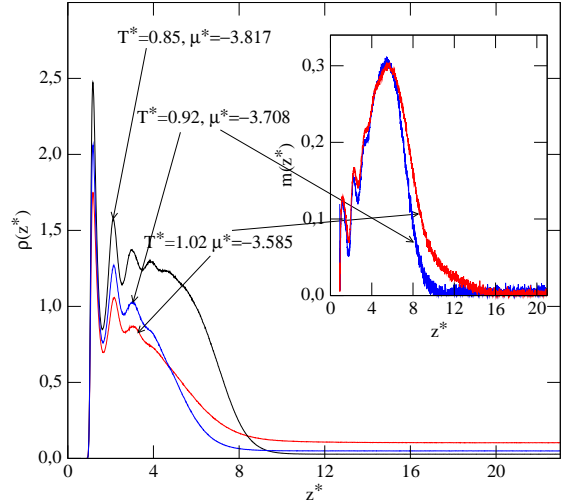}}
\caption{(Colour online) The main figure shows the density profiles for the system with $\text{\egsAA}=6.0$ and $\Delta\varepsilon_{\text{gs}}^{\ast} = 0.2$ at different 
temperatures and at different values of the chemical potential (given
in the figure), while the inset gives the corresponding order parameter profiles. }
\label{fig10}
\end{figure}

Figure~\ref{fig10} presents the density and the order parameter profiles at different temperatures, recorded close to the bulk coexistence. At the temperatures $T^{\ast} = 0.85$, well
below the triple point $T^{\ast}_{\text{tr,v-ml-dl}}$,
the film attains a considerably higher density and thickness than at $T^{\ast}\geqslant T^{\ast}_{\text{tr,v-ml-dl}}$. This becomes understandable by considering that in the mixed film the AB pairs are preferred, 
and a rather small value of the parameter~$s$ allows more molecules to be present in the adsorbed film. One should note that the ratio of densities in the perfectly mixed and demixed thick films
is roughly proportional to $s^{-3}$. 
This relation results from the fact that the average distance between the nearest neighbors in the mixed and demixed states is proportional to $s$.

Above the triple point temperature, the film shows a gradual demixing upon approaching the bulk coexistence. Of course, the observed degree of demixing 
decreases when the temperature grows, and this can be associated with the increasing entropic effects
and with the widening of the interface between the film and the vapour (see figure~\ref{fig10}).  

A qualitatively similar behaviour has been observed for the higher value of $\Delta\varepsilon_{\text{gs}}^{\ast}=0.4$. The wetting temperature has been
found to be lower and equal to $T^{\ast}_{\text w}\approx 0.825$. However, the demixed film of moderate thickness and density is stable only at $T^{\ast}$ above 0.84. At lower temperatures, the  
transition between  thin and thick films leads to the formation of a mixed film of high density. At higher temperatures, a demixed film exists over a certain range of the chemical potential. Only close to 
the bulk coexistence, a reentrant mixing occurs. 

\begin{figure}[!t]
\centerline{\includegraphics[scale=0.4]{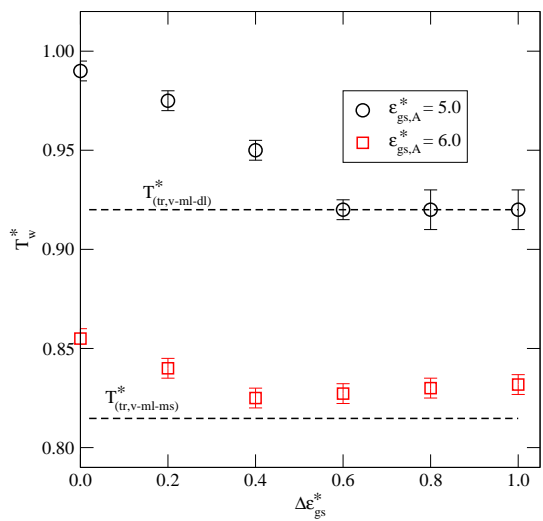}}
\caption{(Colour online) The changes of the wetting temperature against the difference between the adsorption energies of the components
($\Delta\varepsilon_{\text{gs}}^{\ast}$) for the systems with $s=0.73$.}
\label{fig12}
\end{figure}

A further increase of $\Delta\varepsilon_{\text{gs}}^{\ast}$ to 0.6, 0.8 and 1.0, leads to a rather unexpected behaviour. A large difference between {\egsAA} and {\egsBB} causes a demixed film, which is formed 
via the layering transition, to remain stable up to the bulk coexistence over a certain range of temperatures above the triple point $T^{\ast}_{\text{tr,v-ml-ms}}\approx 0.815$. Such a film cannot 
wet the surface, since the coexisting bulk liquid is mixed. The reentrant mixing transition in the film occurs only when the temperature becomes high enough. 
In the case of $\Delta\varepsilon_{\text{gs}}^{\ast}=0.6$, 0.8 and 1.0, it happens only when $T^{\ast}$ is about 0.825, 0.828 and 0.833, respectively (see figure~\ref{fig12}). 

The above results suggest that the complete wetting below the triple point $T^{\ast}_{\text{tr,v-ml-dl}}$, occurs only over the range of temperature 
when the reentrant mixing takes place in the film. One can explain the observed changes in the 
wetting temperature by the following argument. When the {\egsBB} increases, the adsorbed particles experience a stronger attraction to the surface and the film shows an increasing 
demixing tendency. A demixing in the film is also enhanced by the weak AB interaction. These two energetic effects are responsible for the formation of a rather thick demixed layer at low temperatures. 
Such demixed films retain stability up to the bulk coexistence.

Upon a gradual increase of the temperature, the packing effects, which favour the formation of high density mixed films, become stronger and lead to  
a reentrant mixing transition.  
Therefore, the onset of a reentrant mixing is shifted to higher temperatures, leading to an increase of the wetting temperature as well. In fact, when $\Delta\varepsilon_{\text{gs}}^{\ast}=1.4$,
the demixed films have been observed at the temperatures up to about $T^{\ast}=0.84$.

\begin{figure}[!t]
\centerline{\includegraphics[scale=0.4]{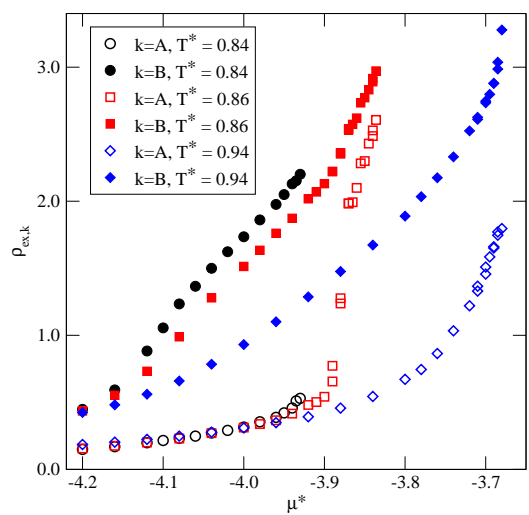}}
\caption{(Colour online) The examples of the excess adsorption isotherms of the components A and B at different temperatures (given in the figure) obtained for the system with $\text{\egsAA}=7.0$
and $\Delta\varepsilon_{\text{gs}}^{\ast} = 1.4$.}
\label{fig11}
\end{figure}

We have performed an additional simulation for $\text{\egsAA}=7.0$ and assuming different values of $\Delta\varepsilon_{\text{gs}}^{\ast}$ between 0.2 and 1.4. We recall that in the case of a non-selective 
wall, the system with $\text{\egsAA}= \text{\egsBB}= 7.0$, exhibits the triple point wetting at $T^{\ast}_{\text{tr,v-ml-ms}}$. The calculations have shown that when  $\Delta\varepsilon_{\text{gs}}^{\ast}\leqslant 1.0$, a 
reentrant mixing occurs at any temperature between $T^{\ast}_{\text{tr,v-ml-ms}}$ and $T^{\ast}_{\text{tr,v-ml-dl}}$, indicating that the triple point wetting survives.
However, when  $\Delta\varepsilon_{\text{gs}}^{\ast} = 1.4$, the reentrant mixing has not been found at the temperatures slightly above the triple point $T^{\ast}_{\text{tr,v-ml-ms}}$, but it takes place when 
the temperature becomes high enough. Figure~\ref{fig11} presents the examples of the excess adsorption isotherms for the components A and B at $T^{\ast} =0.84$, 0.86 and 0.94,  
which demonstrate that at \ta{0.84} the film remains demixed up to the bulk condensation, while it undergoes a reentrant mixing at \ta{0.86}. At the temperature above the triple point 
$T^{\ast}_{\text{tr,v-ml-dl}}$, the film remains demixed when the chemical potential approaches the bulk coexistence. 


\section{Summary}

We have performed a Monte Carlo study of the wetting behaviour of a strongly non-additive symmetric mixture at selective structureless walls. Here, we have considered the mixture with $e=0.6$
and $s=0.73$, of the bulk phase diagram like the one shown in figure~\ref{fig1}.  
This mixture exhibits a demixing transition only at sufficiently high temperatures, above the triple point  $T_{\text{tr,v-ml-dl}}^{\ast}\approx 0.92$, in which the vapour 
coexists with the mixed and demixed liquids. At the temperatures below $T_{\text{tr,v-ml-dl}}^{\ast}$, and above the triple point  $T_{\text{tr,v-ml-ms}}^{\ast}\approx 0.815$, in which the vapour, the 
mixed liquid and the mixed solid coexist, the vapour condenses into a mixed liquid. 

The summary of our findings is given in figure~\ref{fig12}, which presents the changes of the wetting temperature versus $\Delta\varepsilon_{\text{gs}}^{\ast}$ for 
the two series of systems with $\text{\egsAA}=5.0$ and 6.0. In the case of $\text{\egsAA}=5.0$ and the assumed values of $\Delta\varepsilon_{\text{gs}}^{\ast}$, between 0.2 and 1.0, a complete wetting 
occurs only at the temperatures at which the bulk liquid is demixed. An increase of {\egsBB} enhances a demixing in the adsorbed film and leads to a gradual decrease of the 
wetting temperature. At $\Delta\varepsilon_{\text{gs}}^{\ast}\approx 0.6$, the wetting occurs right at the triple point $T_{\text{tr,v-ml-dl}}^{\ast}$. The triple point wetting \cite{theor3} 
has also been observed for $\Delta\varepsilon_{\text{gs}}^{\ast}=0.8$
and 1.0. The lack of a complete wetting below the triple point $T_{\text{tr,v-ml-dl}}^{\ast}$ when {\egsBB} is large, results from the fact that at $T^{\ast}<T_{\text{tr,v-ml-dl}}^{\ast}$ the adsorbed film remains 
demixed at the bulk coexistence, while the bulk liquid in mixed. 

The system with $\text{\egsAA}=6.0$ has been found to behave differently. In the case of $\text{\egsAA} = \text{\egsBB}$, the wetting transition has been estimated \cite{AP-2} 
to take place well below the triple point $T_{\text{tr,v-ml-dl}}^{\ast}$, at $T^{\ast}_{\text w}\approx 0.855$. However, a complete wetting ceases as soon as the temperature reaches the triple point $T_{\text{tr,v-ml-dl}}^{\ast}$.
This is caused by the lack of demixing in the adsorbed film at the bulk coexistence. Only at the temperatures approaching the tricritical point $T^{\ast}_{\text{trc},2}$, another wetting
transition is expected to occur \cite{wet1,wet2}. When the surface becomes selective with respect to the mixture components ($\Delta\varepsilon_{\text{gs}}^{\ast}>0$), the situation changes considerably.
The preferential adsorption of the component B leads to a partial demixing in the film already in the region of low adsorption. When, the difference between {\egsAA} and {\egsBB} is small, 
the developing film undergoes a transition to the demixed thick film, but another transition to a still thicker mixed film takes place upon  approaching the bulk coexistence. This
mechanism has been found to take place at the temperatures well below the triple point $T_{\text{tr,v-ml-dl}}^{\ast}$. Since the bulk liquid is also mixed, a complete wetting occurs, and the 
wetting temperature decreases when $\Delta\varepsilon_{\text{gs}}^{\ast}$ becomes lager. A decrease of the wetting temperature can be attributed to a gradual increase of the average interaction 
energy between the fluid particles and the surface when {\egsBB} increases. However, when the difference between {\egsAA} and {\egsBB} becomes large enough, the demixed film has  not been observed to
undergo the reentrant mixing transition at sufficiently low temperatures.

In the case of non-selective walls, the increase of {\egsAA}  has lowered the wetting temperature \cite{AP-2}, leading a 
triple point wetting at $T^{\ast}_{\text w} = T^{\ast}_{\text{tr,v-ml-ms}}$, i.e., to the temperature at which the vapour coexists with a mixed liquid and solid phases. For the case of 
selective walls, the lack of a reentrant
mixing at the temperatures above the triple point $T^{\ast}_{\text{tr,v-ml-ms}}$ causes  the wetting temperature not to reach the triple point, 
and $T^{\ast}_{\text w}$ to stay above $T^{\ast}_{\text{tr,v-ml-ms}}$. When the difference between 
{\egsAA} and {\egsBB} increases, the stability of the demixed films also increases and this causes a gradual increase of the wetting  temperature.

The scenarios described above may change when the parameters $e$ and $s$ are changed, since the bulk mixtures may show a different phase behaviour \cite{AP-NEW}, 
and therefore, such mixtures may also show different wetting properties. It should be emphasized that we are not aware of any real 
symmetric mixtures that show the bulk behaviour as the one considered in this work. Therefore, it is not possible to compare the results
with any experimental data. However, it might be of interest to study the wetting behaviour of associating mixtures \cite{wetmix8,BOR}, which are known to often
exhibit closed immiscibility loops \cite{LOOPS}.

\ukrainianpart
\title{Неадитивні симетричні суміші на селективних стінках}
\author{A. Патрикєєв}
\address{Відділення моделювання фізико-хімічних процесів,  хімічний факультет, Університет Марії Кюрі-Склодовської, 20031 Люблін, Польща}

\makeukrtitle

\begin{abstract}
Описано результати Монте Карло симуляцій для адсорбції та поведінки змочування високо неадитивної симетричної суміші на селективних стінках. Увага сфокусована на  взаємозв’язку між поверхнево індукованим незмішуванням в адсорбованих плівках і властивостями об'ємної суміші, яка демонструє замкнуту петлю незмішування. Показано, що поведінка змочування залежить від абсолютних значень параметрів, що визначають сили взаємодії між компонентами суміші та поверхнею, а також від їх різниці. Загалом, збільшення різниці між енергіями адсорбції компонент приводить до зменшення температури змочування.
У випадках, коли змочування неселективних стінок відбувається при температурах вищих, ніж настання переходу незмішування в об'ємі, збільшення селективності стінок приводить до поступового зменшення температури змочування до потрійної точки, в якій пара співіснує зі змішаною та незмішаною рідкими фазами. Коли температура змочування на неселективній стінці знаходиться нижче настання переходу незмішування  в об'ємній суміші, збільшення енергії адсорбції вибраного компонента змушує  адсорбовані плівки, що утворюються, незмішуватися та проявляти реентрантне змішування після наближення до    об'ємного співіснування. При температурах вище точки  настання переходу незмішування в об'ємі, адсорбовані плівки залишаються незмішаними аж до об'ємного співіснування та піддаються переходові змочування першого роду. Доволі несподіваною знахідкою виявилось спостереження поступового збільшення температури змочування на високо селективних стінках.

\keywords змочування, методи Монте Карло, поверхнево індуковане фазове розшарування

\end{abstract}

\end{document}